\documentstyle [aps,prl,floats,epsf,epsfig]{revtex}

\begin{document}
\draft
\preprint{HEP/123-qed}
\newcommand{\dm}       {\Delta m^2}
\newcommand{\sinq}      {sin^2 2\theta}

\wideabs{
\title{Predicting Neutron Production from Cosmic-ray Muons}

\author{Y-F.~Wang, V.~Balic, G.~Gratta }
\address{Physics Department, Stanford University, Stanford CA 94305, USA}

\author{A. Fass\`o, S. Roesler }
\address{Stanford Linear Accelerator Center, Stanford CA 94309, USA}

\author{A. Ferrari }
\address{INFN, Sezione di Milano, Via Celoria 16, I-20133, Milano, Italy}

\date{\today}
\maketitle
\begin{abstract}

Fast neutrons from cosmic-ray muons are an important background to underground 
low energy experiments.   The estimate of such background is often
hampered
by the difficulty of measuring and calculating neutron production with
sufficient
accuracy.   Indeed substantial disagreement exists between the different
analytical
calculations performed so far, while data reported by different
experiments is not
always consistent.   We discuss a new unified approach to estimate the
neutron 
yield, the energy spectrum, the multiplicity and the angular distribution 
from cosmic muons using the Monte
Carlo 
simulation package FLUKA and show that it gives a good description of most
of the 
existing measurements once the appropriate corrections have
been applied.
\end{abstract}


}

\narrowtext

\section{introduction}

Fast neutrons from cosmic-ray muons represent an important background 
for low-energy underground experiments such as searches for proton decay 
and dark matter, and low-energy neutrino oscillation.
Unlike charged hadrons which can be tagged by a veto detector system, 
neutrons usually cannot be identified until they are finally captured, 
mimicking the signal. The occasional neutron scattering with the 
sensitive material of the detector and the long lifetime of neutrons 
in the detector and surrounding materials further complicate the 
situation.   For example, the Palo Verde reactor neutrino oscillation 
experiment found such neutrons to be their dominant
background~\cite{palo}, and a similar situation is expected at the ultra-long
baseline detector KamLAND~\cite{KamLand}. 
Low-energy solar neutrino experiments such SNO and Borexino also have to estimate
such backgrounds and the understanding of neutron backgrounds may 
be relevant in resolving 
the controversy between the CDMS~\cite{cdms} and DAMA~\cite{dama} 
results on dark matter searches. Finally, low energy accelerator experiments
at shallow depths, such as Karmen, LSND and OrLaND have also similar problems.

Although the total neutron yield from cosmic muon spallation has been
measured by several experiments, contradictory results are given in 
the literature~\cite{mea,paloneu,chen,lvd,Allk}.
Theoretical calculations~\cite{Allk,petr,Rya} are also not consistent 
with each other and with data. The fact that primary neutrons, pions 
and protons can all produce secondary neutrons through hadronic interactions 
makes analytical calculations very difficult. 
A simple cascade model~\cite{Rya} suggests that the number of nuclear 
cascade products
such as neutrons, pions and protons, increases with the average muon energy
approximately like $\mathrm E_\mu^{0.7}$.
This formula 
agrees with most measurements~\cite{mea,paloneu,chen} except the
recent one by the LVD collaboration~\cite{lvd}.

In addition to these problems with neutron yields, the few measurements 
of the neutron energy spectrum~\cite{lvd,karmen} are not well reproduced by
theoretical calculations~\cite{barton,perkins}.
The interpretation of experimental data is complicated by the fact that
the 
neutron energy spectrum depends upon the muon spectrum that, in turn, is a 
function of the depth at which the measurement is carried-on and the 
geometrical configuration of the underground site.  There is a broad range 
of results reported in the literature.   Barton~\cite{barton} suggests that 
the spectrum of neutrons from hadronic cascade follows $E^{-1/2}$ between 
10-50 MeV, while the spectrum of neutrons from $\pi^-$ capture follows a 
flat spectrum up to 100 MeV.   Perkins~\cite{perkins} suggests that the 
neutron spectrum from muon spallation follows $ E^{-1.6}$. The combination 
of $ (9.7E^{-1/2}+6.0e^{-E/10})~$ has been used in a
measurement~\cite{chen} 
at a shallow site.   It has also been suggested~\cite{wang} to use
proton 
and neutron spectra following $ E^{-1.86}$ as  measured at accelerators 
for photo-nuclear interactions~\cite{khalchukov}. Experimentally the
Karmen 
experiment reported a visible energy spectrum following $ e^{-E/39}$ 
for spallation neutrons~\cite{karmen} and the LVD experiment 
reported a visible energy spectrum following $E^{-1}$.

In order to put some order in this area, we have studied the neutron 
yield, the neutron energy spectrum, the multiplicity   and the angular
distribution using the 1999 version of the 
FLUKA~\cite{fluka} Monte Carlo program which is expected to give an 
accurate description of all the processes involved.   It is our intention 
to provide an unified approach and obtain a reliable estimate of neutron 
background for experiments at all depths. 
While we perform the simulation in a particular detector configuration, 
results should be applicable to a range of substances, including rock and other
shielding materials.

\section{Muon spallation models}

Fast neutrons from cosmic-ray muons are produced in the following
processes:
\begin{itemize}
\item [a)] Muon interactions with nuclei via a virtual photon producing a
nuclear
           disintegration. This process is usually referred to as ``muon 
           spallation'' and is the main source of theoretical uncertainty. 
\item [b)] Muon elastic scattering with neutrons bound in nuclei.
\item [c)] Photo-nuclear reactions associated with electromagnetic showers 
           generated by muons.
\item [d)] Secondary neutron production following any of the above processes.
\end{itemize}

Processes b) and c) are reasonably well understood while a) and d) are the 
root of the difficulties described in previous calculations.    
Neutrons can be also produced from
muons 
which stop and are captured, resulting in highly excited isotopes emitting  
one or more neutrons.    This process is reasonably well understood and
its contribution to total neutron yield can be calculated.  All the
experimental 
results referred to in this paper do not include these neutrons since they can 
be easily identified and eliminated. Neutrons produced from neutrinos 
are negligibly small at the depths considered, and thus are not discussed in this 
paper.

\begin{figure}[hbt!]
\begin{center}
\mbox{\epsfig{file=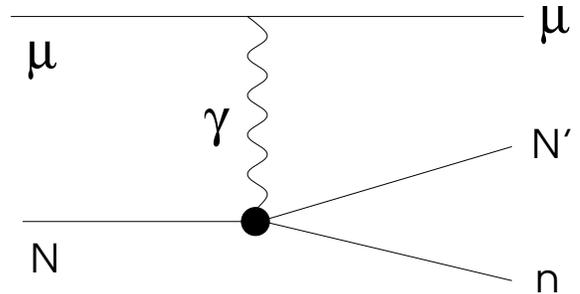,height=4.cm}}
\vskip 0.1cm
\caption{The Feynman diagram of a muon spallation process.}
\label{fig:feynman}
\end{center}
\end{figure}

The muon spallation process is schematically illustrated in
Figure~\ref{fig:feynman}. 
The desired $\mu-N$ cross section is then 
calculated as 
\begin{equation}
 \sigma_{\mu-N} = \int {N_{\gamma}(\nu)\sigma^{virt}_{\gamma-N}(\nu)
                  \over \nu}d\nu
\end{equation}
where $\mathrm \nu=E-E'$, $\mathrm E$ and $\mathrm E'$ are energies of 
initial and final
muons, and $\mathrm N_{\gamma}(\nu)$ the virtual photon energy spectrum.
Theoretical calculations often treat the virtual photons according to the 
Weizs\"acker-Williams approximation~\cite{Jackson}, in which the passage of
a charged particle in a slab of material produces the same effects as of a 
beam of quasi-real photons. A general expression of the
Weizs\"acker-Williams formula 
is given in ref.~\cite{petr}:
\begin{eqnarray}
N_{\gamma}(\nu) & = &{\alpha\over\pi}[ {E^2+E'^2\over p^2}
\ln {EE'+PP'-m^2\over m\nu} \nonumber \\
&-& {(E+E')^2\over 2P^2}\ln {(P+P')^2\over (E+E')\nu}-{P'\over P}]
\end{eqnarray}
where $\mathrm m$ is the muon mass, and $\mathrm P$ and  $\mathrm P'$ 
are momentum of initial and final
muons.

Since in the above approximation, it is assumed that the $\gamma-N$ cross 
section is the same for real and virtual photons, the measured 
$\gamma-N$ cross section can be used to calculate the $\mu-N$ cross 
section in Eq.(1).   
  At low muon energy the situation is more complicated. Here, the virtuality
  of the photon becomes comparable to its energy and cannot be neglected. It
  follows that the Weizs\"acker-Williams approximation can no longer be used.
  In addition, the interaction of the virtual photon with the nucleus is a
  collective excitation of the nucleus (Giant Dipole Resonance, GDR) rather
  than a single photon-nucleon interaction. This implies that the GDR model
  would have to be applied to virtual photons introducing further
  theoretical and technical complications.
  However, it might be reasonable to assume that neutron production by 
low-energy muon interactions is small as compared to neutron photoproduction
  by low-energy bremsstrahlung photons and adds therefore only a minor
  contribution to the total neutron yield.


In addtional to these assumptions,
there are a number of problems associated with
analytical 
calculations: first, they cannot reliably calculate all daughter
products 
for every nucleus if the $\gamma-N$ interaction is very violent so that the
nucleus becomes highly excited;
second, they cannot properly take into account secondary neutron
production.
Hence, while these calculations provide useful guidance and, at shallow
depths, 
where hadronic shower effects are small, they may even give quantitatively
sound
predictions~\cite{petr}, in general they cannot be considered particularly 
reliable.

Monte Carlo approaches are commonly used to properly model hadronic
cascades.
Currently the most complete code to describe both
hadronic and electromagnetic interactions up to
20~TeV is FLUKA~\cite{fluka}.
In this program, 
different physical models, or event generators, are responsible
  for the various aspects of particle production at different energies~\cite{flukamodel}.
  High-energy hadronic interactions are described based on the Dual Parton
  Model followed by a pre-equilibrium-cascade model. In addition, models for
  nuclear evaporation, break-up of excited fragments and $\gamma$-deexcitation
  treat the disintegration of excited nuclei. Hadronic interactions of photons
  are simulated in detail from threshold (Giant Dipole Resonance interactions)
  up to TeV-energies (Vector Meson Dominance Model). 
For nuclei up to copper, measured photo-nuclear 
cross sections in the low energy region are used~\cite{photo}.  
Hadronic interactions of
  muons are based on the Weizs\"acker-Williams approximation as formulated
  by Bezrukov and Bugaev~\cite{Bezrukov}. A spectrum of virtual photons is generated
  which interact with the nuclei similar to real photons. Due to the above
  theoretical and technical complications in the description of hadronic
  interactions of virtual photons at very low energies the simulation is
  restricted to photon energies above the delta resonance threshold.
  The implementation of hadronic interactions of muons has been shown to
  give reliable predictions for the MACRO experiment~\cite{macro}.


In the following FLUKA is used to obtain a consistent and complete
estimate of
neutron production from cosmic muons.   We model a simple cubic detector
filled 
with liquid scintillator $C_nH_{2n+2}$, where $n$ is taken to be 10. Muons
with 
monochromatic energy are tracked in the detector and secondary neutrons
and other 
hadrons are analyzed.

\section{Neutron yield}

The total neutron yield is probably the most measured quantity in our 
problem. There are many experimental results from different depths which 
can be compared with the model.

\begin{figure}[hbt!]
\begin{center}
\mbox{\epsfig{file=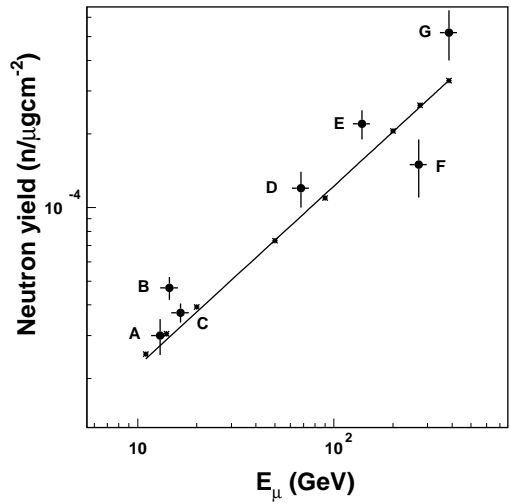,height=8cm}}
\caption{
Neutron production rate as a function of muon energy. 
The stars indicate the FLUKA simulation 
with a fit to the power law.
The experimental points, with abscissa corresponding to the average energy
at the experiment's depth: 
A) 20 meter water equivalent (m.w.e.)~\protect\cite{chen,paloneu},
B) 25 m.w.e.~\protect\cite{bezru}, 
C) 32 m.w.e. by the Palo Verde experiment~\protect\cite{paloneu}, 
D) 316 m.w.e.~\protect\cite{bezru},
E) 750 m.w.e.~\protect\cite{enikeev}, 
F) 3650 m.w.e. by the LVD experiment at Gran Sasso~\protect\cite{lvd}, 
and G) 5200 m.w.e. by the LSD detector at Mont Blanc~\protect\cite{mea}.
}
\label{rate_e}
\end{center}
\end{figure}

All neutrons, either primary or secondary, are included.  Double counting 
due to neutron scattering or neutron spallation is carefully avoided. Due
to 
the limited size of the detection volume in the simulation, some neutrons 
can escape, resulting in fewer secondary neutrons. The total number of 
neutrons thus depends on the size of the detector. 
The detector size has to be limited so that the muon energy-loss is small and
the initial muon energy can be used as a constant. 
To correct 
this problem, we run the simulation with different detector sizes at each 
muon energy, and fit the neutron yield as a function of percentage of 
neutrons which escape the detector. An exponential behavior is found and 
the total number of neutrons can be extrapolated.   The corrected neutron 
yield as a function of the muon energy is shown in Fig.~\ref{rate_e}. 
The neutron yield per muon can be fit as
\begin{equation}
{\mathrm N_n = 4.14\times E^{0.74}_{\mu} \times~ 10^{-6} ~neutron/(\mu g cm^{-2})}
\end{equation}
where E is in GeV. This relation is 
consistent with $\mathrm E^{0.7}_{\mu}$ law 
suggested in ref.~\cite{Rya}. 

While many experiments report their results as a function of the
detector's
depth underground, clearly the proper physical parameter is the mean muon 
energy at the detector.   The conversion between the two quantities is not
entirely trivial as the average energy depends upon the geometry of the 
overburden, particularly in the case of deep sites.    Here for
consistency
we report all experimental results as a function of the average 
muon energy $\mathrm \bar E_{\mu}$.    
For measurements performed at deep laboratories such as Gran Sasso or Mont
Blanc the conversion is given in the original papers~\cite{mea,lvd},
while for shallow sites we estimate the average muon energy-loss using the 
simple relation~\cite{gaisser} and assuming a flat geometry:
\begin{equation}
{dE_{\mu}\over dX} = -\alpha - E_{\mu}/\xi  
\end{equation}
where $\mathrm E_{\mu}$ is in GeV, 
$\mathrm\alpha=(1.9 + 0.08~log(E_{\mu}/m_\mu))\times 10^{-3} ~GeV/(gcm^{-2})$ 
and $\mathrm\xi= 2.5\times 10^5 ~g/cm^2$ in rock.
All known 
experimental neutron-production rates are reported in Figure~\ref{rate_e} along with
the FLUKA predictions.   The agreement is substantially better than
obtained
with previous calculations~\cite{Allk,petr}.

\begin{figure}[hbtp]
\begin{center}
\mbox{\epsfig{file=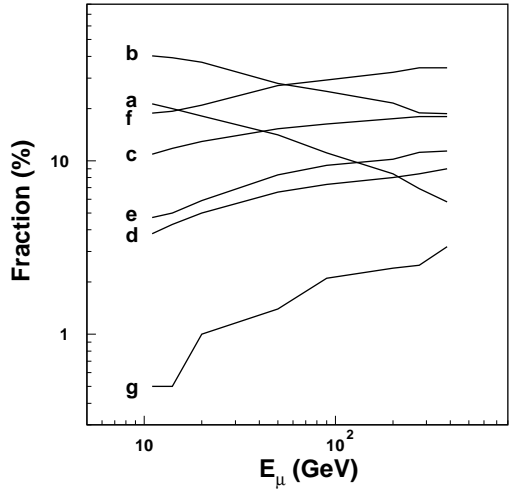,height=8cm}}
\caption{Origin of neutrons: 
a) direct muon spallation, 
b) real photo-nuclear disintegration,
c) neutron spallation,
d) proton spallation,
e) $\pi^+$ spallation,
f) $\pi^-$ spallation and capture,
g) others.
}
\label{fig:source}
\end{center}
\end{figure}

Given this agreement  we can proceed to extract from the 
Monte Carlo simulation information on the origin of the neutrons.  This is of great 
interest as it can provide hints for a better theoretical understanding
of the processes involved.   In Figure~\ref{fig:source} we show the fractional
composition 
of the neutrons by origin.   As expected the fraction of neutrons from 
primary processes such as muon spallation and real photo-nuclear
interactions 
decreases with energy, while secondary neutrons from neutron and proton 
spallation, from pion absorption and from other minor processes such as 
$\Lambda$ and $\Sigma$ decays, increase in relative importance with energy.

The $\pi^+$ yield in muon spallation can be estimated from analytical 
calculations more reliably than that of neutrons. This is because 
1) secondary processes for neutrons are more important;
2) there is an uncertainty in the direct neutron production from 
virtual or real pions.
Table~\ref{tab:pi} shows the comparison of our simulation with an
analytical 
calculation~\cite{petr} which does not include
secondary pions from hadronic showers. In the experimental
data~\cite{chen}, $\pi^+$'s are identified through 
their $\rm\pi^+\to\mu^+\to e^+$ decays, and
hadronic shower effects are not corrected for.
In  the pion case, good agreement 
between data, FLUKA and the analytical calculation can be 
seen at shallow depth, where
secondary hadrons from showers are not important.
We obtained the pion yield per muon as:
\begin{eqnarray}
{\mathrm N_{\pi^+} = 4.45\times E^{0.80}_{\mu} \times 10^{-7} ~pion/(\mu g cm^{-2})} 
\end{eqnarray}
where $\mathrm E_{mu}$ is in GeV.  
The $\mathrm E^{0.7}_{\mu}$ law appears to be universal as suggested by 
Ryazhskaya~\cite{Rya} and consistent with results from ref.~\cite{enikeev}.

\begin{table}[htbp!]
\begin{center}
\begin{tabular}{|c|c|c|c|c|c|c|}
   Depth(m)   & \multicolumn{2}{c|}{20} & \multicolumn{2}{c|}{100}  & 
                   \multicolumn{2}{c|}{500} \\
  Energy(GeV) & \multicolumn{2}{c|}{10.3} & \multicolumn{2}{c|}{22.4} & 
                   \multicolumn{2}{c|}{80.0} \\
\hline
  yield       & n & $\pi^+$ & n & $\pi^+$ & n & $\pi^+$ \\
\hline
  anal. calc.~\cite{petr} &  0.87 & 0.30 & 1.21 & 0.45 & 2.08& 0.86  \\
  FLUKA       & 2.5& 0.31  & 3.9 & 0.52 & 11.0 & 1.51 \\
 Data~\cite{chen}  & 3.0$\pm$ 0.5 & 0.35$\pm$0.07& - & - & - & -\\
\end{tabular}
\caption{Neutron and $\pi^+$ yields (in units of $\rm 10^{-5} /(gcm^{-2}$)) per muon 
at different depths. Note that the analytical calculation~\protect\cite{petr} 
does not include real photon-nuclear 
disintegration and secondary particles.}
\label{tab:pi}
\end{center}
\end{table}

\begin{figure*}[hbtp!]
\begin{center}
\mbox{\epsfig{file=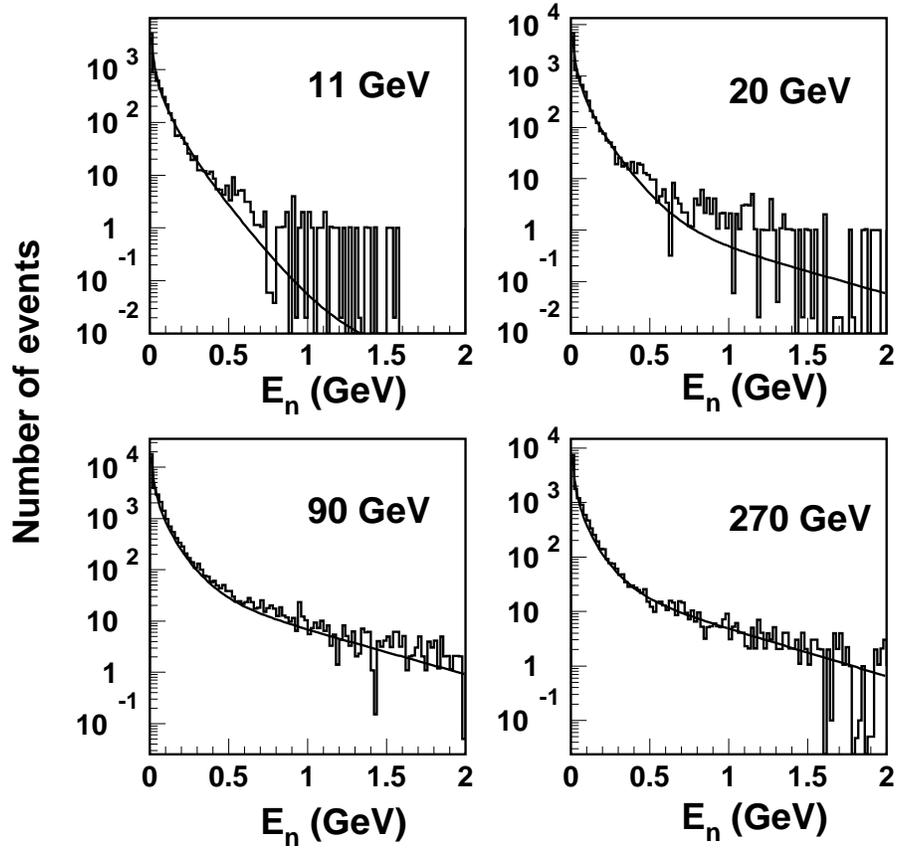,height=14cm}}
\caption{Energy spectrum of neutrons at different muon energies
together with our parameterization. We find  $\chi^2/NDF$ of 
 3.9, 4.5, 9.6 and 3.7 for 11, 20, 90 and 270 GeV respectively.}
\label{fig:energy}
\end{center}
\end{figure*}

\section{Neutron energy spectrum}

The neutron energy spectrum is particularly controversial, with
a wide range of results reported in theoretical calculations and in
the few experimental measurements.   In Figure~\ref{fig:energy} we show some of the 
energy spectra obtained with FLUKA.    Each histogram is fitted to the
universal empirical function 
\begin{equation}
{dN\over dE_n} = A\left({e^{-7E_n}\over E_n}+B(E_\mu)e^{-2E_n}\right)
\end{equation}
where $A$ is a normalization factor, and 
\begin{equation}
B(E_\mu) = 0.52 - 0.58e^{-0.0099E_\mu}.
\end{equation}
This simple function reproduces fairly well the FLUKA distributions with 
 $\chi^2$ per degree of freedom of
3.9, 4.5, 9.6 and 3.7 for 11, 20, 90 and 270 GeV respectively.

\begin{figure}[hbtp!]
\begin{center}
\mbox{\epsfig{file=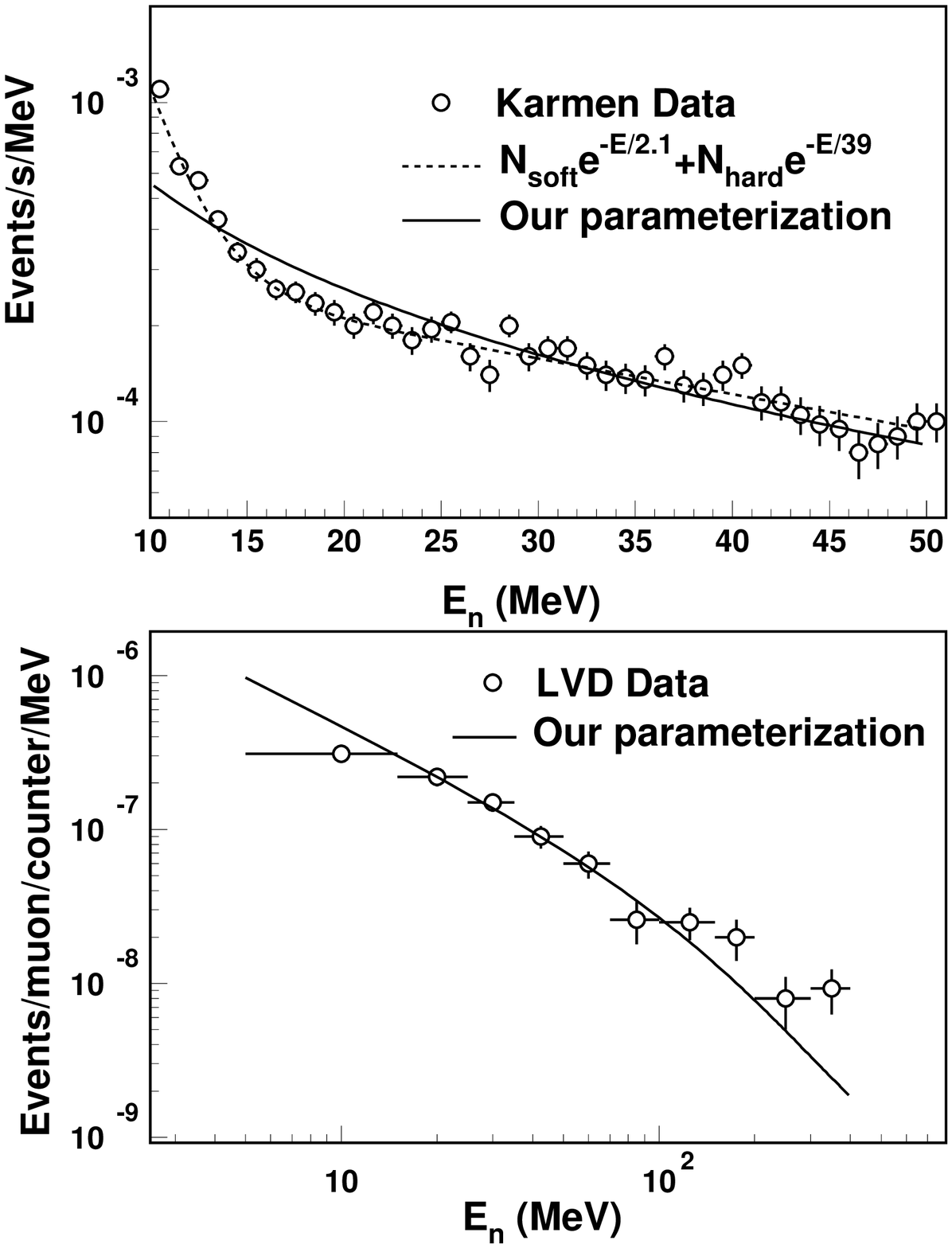,height=10cm}}
\caption{Comparison of measured neutron energy spectrum with our parameterization.}
\label{fig:energy_data}
\end{center}
\end{figure}


We are aware of two direct measurements that can be compared to our
calculations, as shown in Fig.~\ref{fig:energy_data}.  
The Karmen collaboration measured neutron energy spectrum 
up to 50~MeV\cite{karmen}.
Within this modest range, their parameterization 
$\mathrm N_{soft}e^{-E_n/2.1}+N_{hard}e^{-E_n/39}$, 
is in a reasonable agreement 
with our result. They attribute all the soft component to muon capture but
it seems that muon spallation also produces some soft neutrons.
The LVD experiment reported a $E_n^{-1}$
spectrum\cite{lvd} up to 400 MeV, also in a reasonable agreement 
with our result.

Some of the theoretical estimates of neutron energy spectrum are comparable to 
our results at low energies while our calculation gives results over a much wider
range of energies.
For example, the power laws $E^{-1.6}_n$~\cite{perkins} and 
$E^{-1.86}_n$~\cite{wang} agree with our results
up to 400 MeV. Other functions, 
such as  $(9.7E^{-1/2}_n+6.0e^{-E_n/10})~$\cite{chen}, 
or that suggested by Barton~\cite{barton}, are significantly different.

\section{Neutron multiplicity and angular distribution}

\begin{figure*}[hbt!]
\begin{center}
\mbox{\epsfig{file=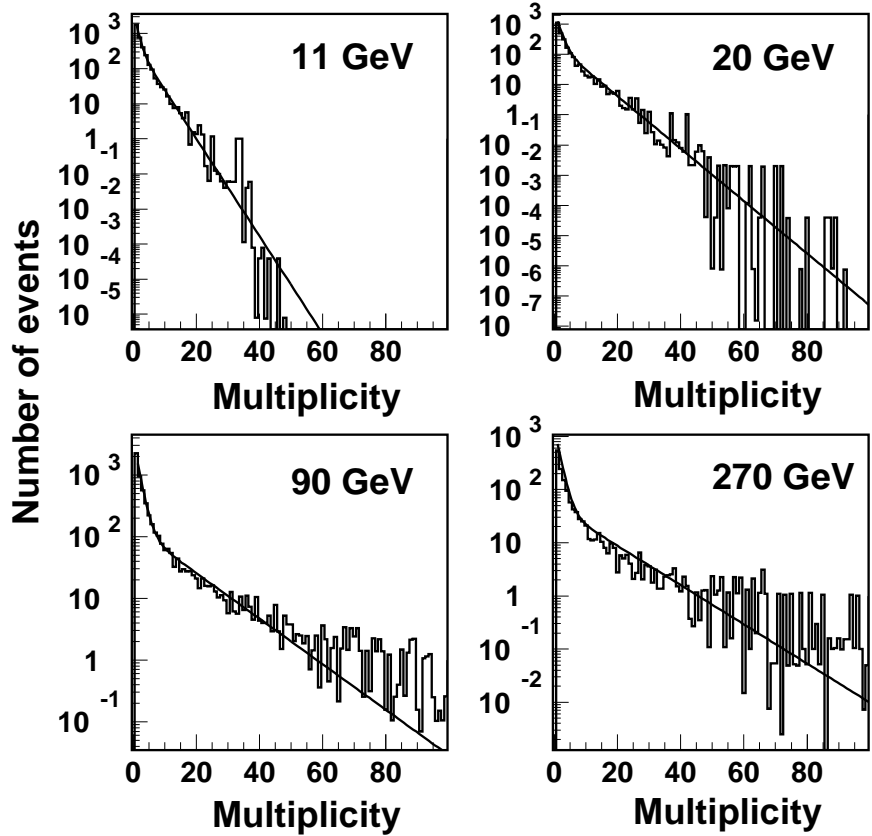,height=14cm}}
\caption{Neutron multiplicity at different muon energies together with 
our parameterization. We find  $\chi^2/NDF$ of 
0.6, 2.0,1.5  and 1.5 for 11, 20, 90 and 270 GeV respectively. }
\label{fig:multi}
\end{center}
\end{figure*}

The neutron multiplicity is probably the least known quantity in the
neutron
production problem.    In most 
cases muon spallation only happens once
and 
produces only a few primary hadrons.  But these hadrons can shower and
generate 
secondary hadrons, including neutrons.   Using our simulation we have
found that
in some cases the number of secondary neutrons exceeds 50.     
The average number of neutrons is about 3 for a 11~GeV muon, and it
increases 
to about 7 at a muon energy of 385~GeV.    Figure~\ref{fig:multi} shows the neutron
multiplicity 
distributions at different muon energies from FLUKA, together with the 
universal empirical parameterization:
\begin{equation}
{dN\over dM} =  A(e^{-A(E_\mu)M}+B(E_\mu)e^{-C(E_\mu)M})
\end{equation}
where $M$ is the multiplicity,
\begin{eqnarray}
A(E_\mu) & =& 0.085+0.54e^{-0.075E_\mu}           \\
B(E_\mu) &=& {27.2\over 1+7.2e^{-0.076E_\mu}}     \\
C(E_\mu) &=& 0.67+1.4e^{-0.12E_\mu}            
\end{eqnarray}
The $\chi^2$'s per degree of freedom are 
0.6, 2.0,1.5  and 1.5 for 11, 20, 90 and 270 GeV respectively.

There are only few experimental results~\cite{paloneu} on multiplicity.  
The Palo Verde experiment at the shallow depth corresponding to
 a mean muon energy of about 
16.5~GeV, observed a two-neutrons to one-neutron ratio  
between 5 to 10, depending on the assumption on three neutron yield. 
Similar results are also reported by 
Bezrukov {\it et al.}~\cite{bezru}. 
These numbers appear to be substantially larger than the FLUKA prediction of 2.
While more data would be desirable to better understand this discrepancy
it is
possible that the data is affected by the incomplete efficiency matrix for
four 
or more neutrons, whose contribution, previously assumed small, seems
significant 
from our simulation.

\begin{figure*}[hbt!]
\begin{center}
\mbox{\epsfig{file=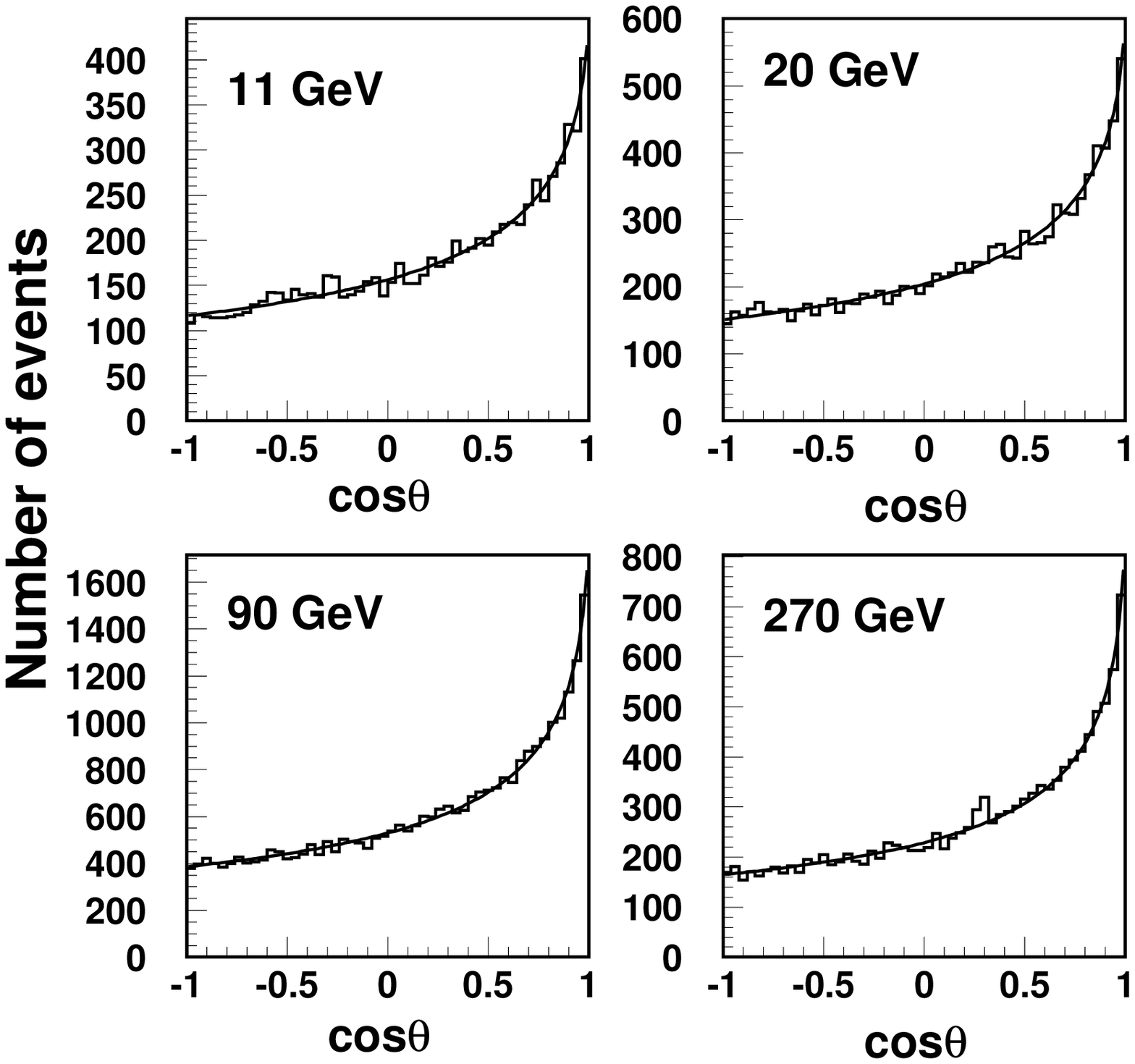,height=14cm}}
\caption{Neutron angular distributions with respect to the muon direction 
at different muon energies together with our parameterization.
We find  $\chi^2/NDF$ of 
0.51, 0.51, 0.72 and 0.72 for 11, 20, 90 and 270 GeV respectively.}
\label{fig:ang}
\end{center}
\end{figure*}

There are no experimental results on the neutron angular distribution
with respect 
to muons. Experimental results~\cite{angular} on 
$\mathrm \gamma + ^{12}C \to p+X$
have been used~\cite{wang} as a first approximation of the neutron
angular 
distribution.    The distribution is expected to be forward-peaked,
smoothed
somewhat by the contribution of secondary neutrons.    Our simulation is
well parameterized by the angular distribution
\begin{equation}
\mathrm {dN\over dcos\theta} = {A \over (1-cos\theta)^{0.6}+{B(E_\mu)}} 
\end{equation}
where $B(E_\mu) = 0.699 E^{-0.136}_\mu$.    Figure~\ref{fig:ang} shows the $\cos\theta$
distribution at the usual set of energies together with the above
parameterization.
The $\chi^2$ per degree of freedom are 
0.51, 0.51, 0.72 and 0.72 for 11, 20, 90 and 270 GeV respectively.

\section{Summary}

We obtained for the first time a complete description of neutron
production 
from cosmic-ray muons using the FLUKA Monte Carlo program. 
Results have been compared with existing data and some analytical
calculations reported in literature.   
With a few exceptions, 
our results agree well with data, and our predictions
cover an energy range much broader than what has been discussed before.

Analytical parameterizations have been obtained 
for neutron yield, energy spectrum, multiplicity and angular distribution. 
These formulae can be used as a starting point to estimate neutron
backgrounds
from cosmic-ray muons when the muon energy spectrum is known. For detailed 
calculation of muon-induced neutron backgrounds in low-energy underground 
experiments, a complete Monte Carlo simulation is needed to account for the 
correlation among different variables and can be setup using FLUKA.
Intermediate processes such as neutron scattering before neutron
spallation
are also automatically taken into account in this method that has great
promise
to improve the quality of predictions.

\section {Acknowledgments}

We would like to thank P. Vogel for many useful discussions.
This work was supported in part by DoE grant DE-FG03-96ER40986 and
DE-AC03-76SF00515.

\end{document}